\documentclass[journal]{aastex61}

\usepackage{floatrow}
\usepackage{hyperref}
\usepackage{natbib}
\shorttitle{Existence of coronal jet precursors in bright points}
\shortauthors{Bagashvili, Shergelashvili, Japaridze et al.}

\begin{document}

\title{Evidence for precursors of the coronal hole jets in solar bright points}
\correspondingauthor{Bidzina M. Shergelashvili}
\email{Bidzina.Shergelashvili@oeaw.ac.at}
\author{Salome R. Bagashvili}
\affiliation{Center for Mathematical Plasma Astrophysics, Department of Mathematics, KU Leuven, 200 B, B-3001,
Leuven, Belgium}
\affiliation{Abastumani Astrophysical Observatory at Ilia State University, Kakutsa Cholokashvili Ave 3/5, 0162
Tbilisi,
Georgia}
\affiliation{Combinatorial Optimization and Decision Support, KU Leuven campus Kortrijk, E. Sabbelaan 53, 8500
Kortrijk, Belgium}
\author{Bidzina M. Shergelashvili}
\affiliation{Space Research Institute, Austrian Academy of Sciences, Schmiedlstrasse 6, 8042 Graz, Austria}
\affiliation{Abastumani Astrophysical Observatory at Ilia State University, Kakutsa Cholokashvili Ave 3/5, 0162
Tbilisi,
Georgia}
\affiliation{Combinatorial Optimization and Decision Support, KU Leuven campus Kortrijk, E. Sabbelaan 53, 8500
Kortrijk, Belgium}
\author{Darejan R. Japaridze}
\affiliation{Abastumani Astrophysical Observatory at Ilia State University, Kakutsa Cholokashvili Ave 3/5, 0162
Tbilisi,
Georgia}
\author{Vasil Kukhianidze}
\affiliation{Abastumani Astrophysical Observatory at Ilia State University, Kakutsa Cholokashvili Ave 3/5, 0162
Tbilisi,
Georgia}
\author{Stefaan Poedts}
\affiliation{Center for Mathematical Plasma Astrophysics, Department of Mathematics, KU Leuven, 200 B, B-3001,
Leuven, Belgium}
\author{Teimuraz V. Zaqarashvili}
\affiliation{Space Research Institute, Austrian Academy of Sciences, Schmiedlstrasse 6, 8042 Graz, Austria}
\affiliation{Institute of Physics, IGAM, University of Graz, Universit\"atsplatz 5, 8010 Graz, Austria }
\affiliation{Abastumani Astrophysical Observatory at Ilia State University, Kakutsa Cholokashvili Ave 3/5, 0162
Tbilisi,
Georgia}
\author{Maxim L. Khodachenko}
\affiliation{Space Research Institute, Austrian Academy of Sciences, Schmiedlstrasse 6, 8042 Graz, Austria}
\author{Patrick De Causmaecker}
\affiliation{Combinatorial Optimization and Decision Support, KU Leuven campus Kortrijk, E. Sabbelaan 53, 8500
Kortrijk, Belgium}
\begin{abstract}
A set of 23 observations of coronal jet events that occurred in coronal bright points has been analyzed. The focus was 
on the temporal evolution of the mean brightness before and during coronal jet events. In the absolute majority of the 
cases either single or recurrent coronal jets were preceded by slight precursor disturbances observed in the mean 
intensity curves. The key conclusion is that we were able to detect quasi-periodical oscillations with 
characteristic periods from sub-minute up to 3-4 min values in the bright point brightness which precede the jets. Our 
basic claim is that along with the conventionally accepted scenario of bright point evolution through new magnetic flux 
emergence and its reconnection with the initial structure of the bright point and the coronal hole, certain MHD 
oscillatory and wave-like motions can be excited and these can take an important place in the observed dynamics. These 
quasi-oscillatory phenomena might play the role of links between different epochs of the coronal jet ignition and 
evolution. They can be an indication of the MHD wave excitation processes due to the system entropy variations, density 
variations or shear flows. It is very likely a sharp outflow velocity transverse gradients at the edges between the open 
and closed field line regions. We suppose that magnetic reconnections can be the source of MHD waves due to impulsive 
generation or rapid temperature variations, and shear flow driven nonmodel MHD wave evolution (self-heating and/or 
overreflection mechanisms).
\end{abstract}


\section{Introduction} \label{sec:intro}
Coronal Jets (CJs) are collimated plasma flows observed in extreme ultraviolet (EUV), X-Rays and white light
images across the entire solar atmosphere: in quite corona and active
regions \citep{Schmieder2013,Chandra2017,Guo2014,Raouafi2016} and in coronal holes (CHs)
\citep{Sterling2015,youngmuglach2014a,youngmuglach2014b,Young2015}.
The contemporary space based observational missions (such as Hinode, Solar Dynamics Observatory, and others) with
state-of-the-art spatiotemporal resolution revealed that jet-like transient processes occur very often in the solar
atmosphere \citep{Savcheva2007,Paraschiv2010}.

\begin{figure*}
\vspace*{1mm}
\begin{center}
\includegraphics[scale=0.7]{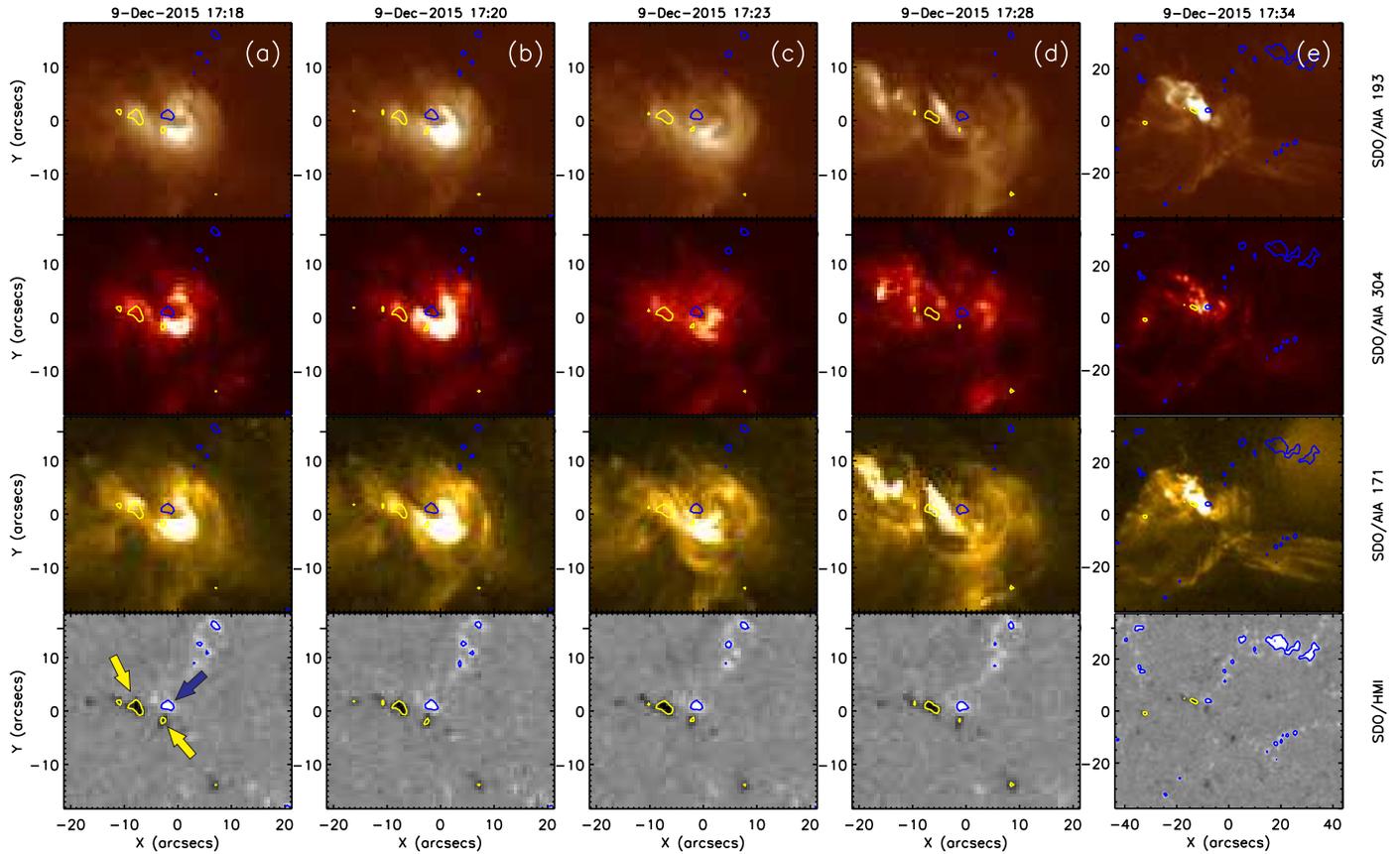}
\end{center}

\caption{ Sample BP before and during CJ ejection process. The top three row represent overlap of \emph{SDO}/AIA 193 {\AA}, 304 {\AA} and 171 {\AA} intensity
images and \emph{SDO}/HMI photospheric magnetograms. The bottom row represents HMI magnetogram separately. Blue and yellow contours indicate positive and negative magnetic field polarities, respectively. These regions are also indicated with arrows in the bottom row of panel (a). Panel (a) - the beginning of precursor; panel (b) - the peak of the precursor brightening; panel (c) - the time after the precursor when still there is no signature of main jet outflow; panel (d) - the moment when the structure is destabilized and the jet-type instability starts; panel (e) - the fully developed transient jet outflow. In panel (e) we take a wider observational window as shown on the corresponding axes.}
\label{Fig1}
\end{figure*}
\begin{figure*}
\vspace*{1mm}
\begin{center}
\includegraphics[scale=0.8]{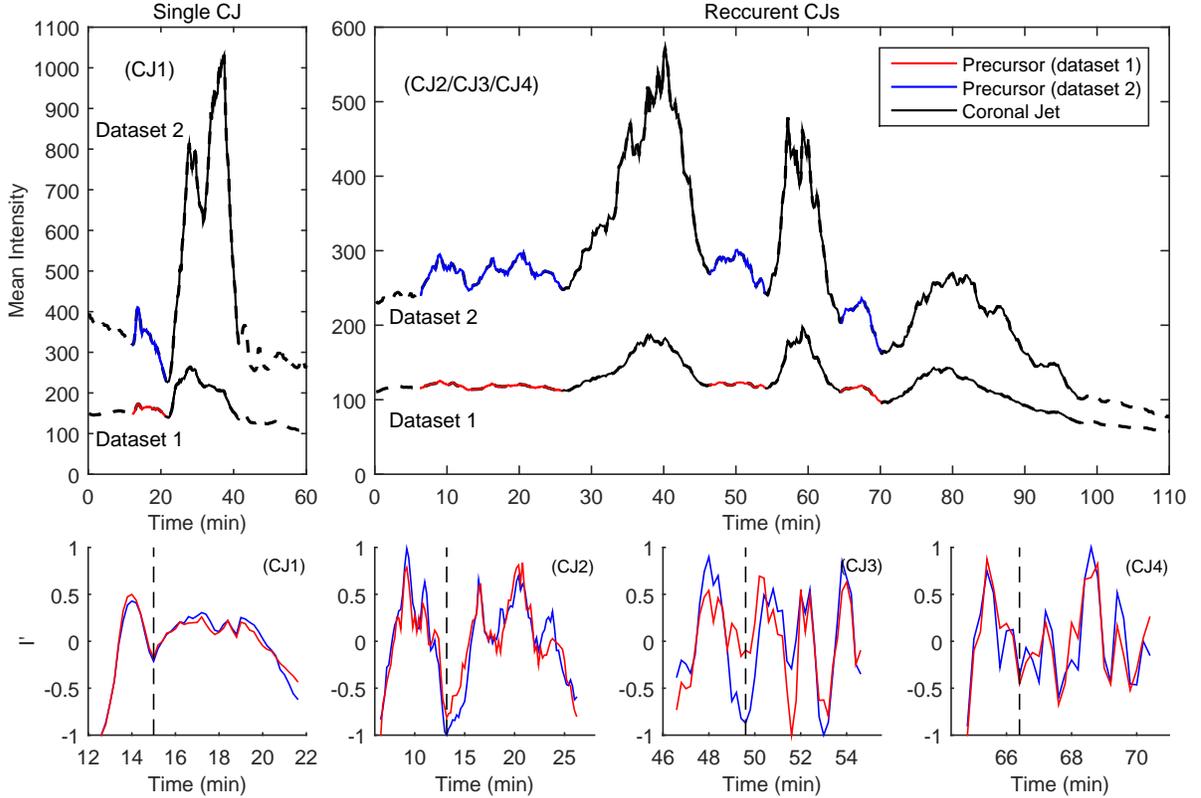}
\end{center}

\caption{Coronal Jet intensity curves for dataset 1 (without threshold - bottom curves) and dataset 2 
with threshold - top curves) in top panels (CJ1) and
(CJ2/CJ3/CJ4). Red (dataset 1) and blue (dataset 2) solid-line parts of the curves indicate precursors events. 
Accordingly, we plot
zooms of these parts in the detrended form in bottom panels (CJ1), (CJ2), (CJ3), (CJ4), respectively. Panels 
(CJ1) (top and bottom) correspond
to the case of the single CJ that started at 2015-12-09 17:28 UT with a precursor start approximately $9.8$ minutes before the CJ
release. While, panels (CJ2/CJ3/CJ4), (CJ2), (CJ3) and (CJ4) demonstrate the case of the recurrent CJs that 
started at 2015-12-30 23:16 UT,
which includes three subsequent plasma ejections. Each of them has precursors starting $19.6$, $7.8$ and $5.6$ minutes before
the respective CJs. Vertical dashed lines represent the end of precursor ignition.}
\label{Fig2}
\end{figure*}
Despite some observational and theoretical studies \citep{Shibata1992, Yokoyama1995, Shimojo2001, Ryutova2008,
Pariat2009, Moore2010, Magara2010, Pariat2010, Archontis2010} the mechanism for CJ formation and evolution is not
entirely understood. There is a general consensus that all of the jet-like features should originate from the common
basic
formation mechanism, viz.\ the release of free magnetic energy due to magnetic reconnection processes
\citep{Shibata2007, Pariat2009}. The majority of the investigated CJs eject from newly emerging or developed bright
points (BPs) \citep{Nistico2009} during the peak of their brightness \citep{Pucci2012}. In the past BPs visible 
in H-alpha and in EUV have been frequently observed before cool jets called surges and collimated with EUV jets 
\citep{Gu1994, Canfield1996, Schmieder1996}. \citet{Sterling2015, Sterling2016} presented new results that suggest that 
mini-filament eruptions
trigger CJs, and there might be no difference in the scenario of standard and blowout jet formation. Furthermore,
\citet{Panesar2016} presented an observational study that reports that mini-filament eruptions, which lead to CJ
acceleration, are driven by  flux cancellation in the photosphere.

The matter of our focus,
in this letter is a particular aspect of the observed dynamics, namely a systematic observational study of
the CJs
associated with on-disk CHs. Our aim was to deduce from the extreme EUV data analysis the character of the temporal behavior of the mean intensity  and the connection to the jet outflow processes themselves.

In general, recent studies report on the BP brightness fluctuations and a variety of their oscillation periods
\citep{Sheeley1979, Nolte1979, Strong1992, Mandrini1996, Aulanier2007, Tian2008, Pontin2011, Kumar2011, Guo2013,
Schmieder2013}. Besides, \citet{Pucci2012} followed the BPs brightness evolution for several CJs and found that most of
the
jets occurred in close temporal association with the brightness maxima. This observational evidence leads us to the
assumption that in the dynamics of BPs, different MHD wavelike or oscillatory disturbances may take place. In this
regard, \citet{Pariat2012} assumes that the observed structure of CJs could be explained using the generation of a
helical kink-like wave. \citet{Lee2015} reports that jets manifest oscillatory motions during their ejection
linking them to wave processes. Moreover, such disturbances may provide a link for the coupling of the dynamic
processes occurring in BPs. It is natural to assume (in the current study only at an intuitive level) that these wavelike
processes can be excited in the substantially non-equilibrium environment such as shear flows \citep{Shergelashvil2006} or
flows with thermal variations \citep{Shergelashvil2007}. This kind of disturbances can be excited in the vicinity of and triggered by the above- mentioned, conventionally accepted coronal jet energy sources viz.\ magnetic reconnection and flux emergence/cancelation processes.  For an analytical modeling under these frameworks, one should
keep in mind that such wave-like dynamical processes follow  more or
less similar scenarios: ignition of a precursor disturbance, its evolution in time, and triggering of a main larger
event. Besides, the overall contribution of CJ incidents in the energy budget of the solar atmosphere and wind can
be estimated using the statistical properties of CHs reported by \citet{Bagashvili2017}, with which the BPs of interest
can be linked. However, before achieving such far reaching consequences regarding analytical modeling, a thorough
observational analysis and confirmation of the regular presence of such quasi-oscillatory type of precursors are needed. These are the
ultimate goals of the observational studies reported in the current letter.
\section{The methodology of observations and data analysis} \label{sec:data}
\begin{deluxetable*}{ccccccccc}[b!]
\tablecaption{Parameters of CJs and their precursors \label{tab:mathmode}}
\tablecolumns{9}
\tablenum{1}
\tablewidth{0pt}
\tablehead{
\colhead{No.} &
\colhead{Obs. start time} &
\colhead{Location} & \colhead{Type} &
\colhead{$\tau_{PI}$} &
\colhead{$\tau_{PT}$} & \colhead{$\tau_{CJ}$} & \colhead{$\Delta \tau_{Peaks}$}  & \colhead{Osc. periods}\\
}
\startdata
CJ1 &	09/12/2015 17:18	& 355, -260	 & 	S	&  1.75	/	1.28	&	 9.4	/	9.8	&	 7.16	 /	 8.13	 &	13.55	/	
14.49	& 1.8 - 3.1 \\
CJ2 &	30/12/2015 22:56	&	-780, -270  & 	R	& 4.58/4.78	&	 19.6/19.6	&	 9.20/10.15	&	 13.83	/	
13.31	& 4.0 - 5.0\\
CJ3 &	30/12/2015 23:40	&	-780, -270  & 	R	& 2.78/2.27	&	 7.8/7.8	&	 4.85/5.54	 &	 8.03	 /	
8.02	& 1.6 - 2.7 \\
CJ4 &	30/12/2015 23:58	&	-780, -270  & 	R	& 2.59/1.51	&	5.6	 /5.6	&	 16.38/16.96	 &	 13.11	/	
12.59	& 0.7 - 2.9\\
CJ5 &	31/12/2015 02:12	&	-770, -280 &	R	& 4.40	/	3.86	&	 13.6	/	 13	&	 11.77	 /	 11.33	&	20.78	 /	
20.95	& 1.4 - 4.6 \\
CJ6 &	31/12/2015 02:52	&	-770, -280 &	R	& 2.97	/	3.05	&	 8	/	8	 &	 6.23	 /	 5.96	 &	14.27	/	
14.14	& 0.9 - 3.4 \\
CJ7 &	07/12/2015 15:56	&	385, 510   &    R	& 13.82	/	12.72	&	 21.2	/	 20	&	 13.49	 /	 15.60	&	20.64	 /	
23.80	& 1.9 - 5.1 \\
CJ8 &	07/12/2015 17:08	&	385, 510   & 	R	& 3.17	/	3.85	&	 4.9	/	 5.4	 &	 8.05	/	 8.24	 &	9.49	/	
8.96	& 1.2 - 2.1 \\
CJ9 &	08/12/2015 11:38	&	15, -225   & 	S	& 4.73  /   5.03	&	 20	/20.2	 &	 6.38	 /	 4.42	 &	19.92	/	
19.43	& 2.3 - 4.1 \\
CJ10 &	05/12/2015 09:11	&	430, 230   & 	S	& 1.72	/	1.62	&	 2.2	/	 2.2	 &	 9.21	/	 8.43	 &	5.71	/	
6.50	& 0.5- 1.2 \\
CJ11 &	10/12/2015 19:25	&	765, -230  &	R	& 5.79	/	6.09	&	 14.6	/	 13.8	&	 5.64	/	 6.50	&	 15.39	/	
15.57	& 1.2 - 3.7 \\
CJ12 &	10/12/2015 21:04	&	765, -235  &	R	& 5.90	/	5.91	&	 7.8	/	 7.8	 &	 10.77	/	 10.52	 &	11.05	/	
11.58	& 1.0 - 4.0 \\
CJ13 &	21/03/2016 21:30	&	515, 585   & 	S	& 12.02	/	12.59	&	 12.59	/	 12.59	&	 12.02	/	 12.59	&	0	 /	
0	& - \\
CJ14 &	21/03/2016 21:21	&	630,  435  &	S	& 6.73	/	6.49	&	 11.6	/	 10.4	&	 11.80	/	 11.51	&	 11.45	/	
12.46	& 0.8 - 1.9 \\
CJ15 &	22/03/2016 02:40	&	-705, -130  & 	R	& 3.57	/	4.05	&	 4.05	/	 4.05	&	 3.57	/	 4.05	&	0	 /	
0	& - \\
CJ16 &	22/03/2016 02:55	&	-660, -80  &	R	& 9.44	/	9.06	&	 9.06	/	 9.06	&	 9.44	/	 9.06	&	0	 /	
0	& - \\
CJ17 &	18/04/2016 10:44	&	-460, 275  &	S	& 1.86	/	1.89	&	 3.2	/	 3.2	 &	 3.36	/	 3.32	 &	3.24	/	
3.15	& 0.4 - 1.7 \\
CJ18 &	18/04/2016 13:39	&	-550, 445  & 	S	& 2.70	/	2.22	&	 4.4	/	 3.4	 &	 10.40	/	 9.96	 &	8.99	/	
10.37	& 0.9 - 1.5 \\
CJ19 &	18/04/2016 14:08	&	-520, 385  &	S	& 0.93	/	0.88	&	 2	/	 2.2	 &	 6.28	/	 5.42	 &	9.12	/	
8.81	& 0.8 - 1.2 \\
CJ20 &	18/04/2016 16:58	&	605, 460   & 	R	& 3.77	/	3.34	&	 4	/	 6.0	 &	 9.87	/	 10.35	 &	17.59	/	
17.05	& 0.5 - 2.1 \\
CJ21 &	18/04/2016 21:25	&	640, 450   &	R	& 1.81	/	2.36	&	 3.2	/	 3.2	&	 6.35	/	 5.99	&	6.40	/	
6.15  & 0.5 - 1.1	\\
CJ22 &	18/04/2016 21:57	&	640, 450   & 	R	& 1.90	/	1.90	&	 2.4	/	 2.4	 &	 6.39	/	 5.81	 &	5.28	/	
5.34	& 0.5 - 0.6 \\
CJ23 &	18/04/2016 18:25	&	-485, 385  &	S	& 3.03	/	3.08	&	 6.2	/	 6.6	 &	 6.12	/	 4.91	 &	8.54	/	
8.27	& 1.8 - 3.6 \\
\enddata
\tablecomments{Complete set of 23 selected CJ and corresponding precursor parameters. Location represents BP (x, y) coordinates in arcsecs; in the column Type we have two possible values S - single and R - recurrent;
$\tau_{PI}$ represents the precursor ignition duration; $\tau_{PT}$ is the precursor total evolution time span; ($\tau_{CJ}$) - CJ durations, $\Delta \tau_{Peaks}$ -  The time intervals between precursor and CJ peaks; Osc. Periods shows the minimum and maximum values of precursor oscillation periods. The parameters having time dimension are measured in minutes.}
\end{deluxetable*}
 \begin{deluxetable*}{cccccc}[b!]
 \tablecaption{Parameters average values CJs and their precursors
\label{tab:parameter_means}}
 \tablecolumns{6}
 \tablenum{2}
 \tablewidth{0pt}
 \tablehead{
 \colhead{Parameters $x$}&\colhead{Maximum probability $f_0$}&\colhead{Expected value $\bar{x}$}&\colhead{ $\sigma$
(variance)}&\colhead{Error$=\sigma /\sqrt{N}$}\\
}
 \startdata
$\tau_{PI}$ &0.21/0.22&2.77/2.85&1.35/1.29&$\pm0.28$/$\pm0.27$\\
$\tau_{PT}$ &0.13/0.15&5.95/5.92&4.26/4.09&$\pm0.89$/$\pm0.85$\\
$\Delta \tau_{Peaks}$ &0.12/0.11&10.25/10.13&6.83/7.59&$\pm 1.43$/$\pm 1.58$\\
$\tau_{CJ}$ &0.18/0.18&8.52/8.62&4.03/4.28&$\pm 0.84$/$\pm 0.89$\\
\hline
$I_{CJ}/I_P$ &0.31/0.28&2.18/2.59&0.80/1.12&$\pm0.17$/$\pm0.23$\\
$\tau_{CJ} /\tau_{PI}$ &0.17/0.19&2.02/1.69&1.16/0.95&$\pm 0.24$/$\pm 0.20$\\
$\Delta \tau_{Peaks}/\tau_{PI}$ &0.23/0.20&2.72/2.56&1.38/1.69&$\pm 0.29$/$\pm 0.35$\\
$\Delta \tau_{Peaks}/\tau_{CJ}$ &0.16/0.16&1.36/1.34&0.67/0.67&$\pm 0.14$/$\pm 0.14$\\
\hline
Min. osc. period & - & $1.24$ & $ 0.83$ & $\pm 0.18$ \\
Max. osc. period & - & $2.77$ & $1.34$& $\pm 0.30$\\
\enddata
\tablecomments{ The CJ and corresponding precursor parameters with related variance and error estimations. $\tau_{PI}$ represents the precursor ignition duration; $\tau_{PT}$ is the precursor total evolution time span;
($\tau_{CJ}$) - CJ durations, $\Delta \tau_{Peaks}$ -  The time intervals between precursor and CJ peaks. All the quantities in the top four rows are measured
in minutes and in bottom four rows all are dimensionless. All values are given in accordance with the order data set 1/data set 2. In two bottom row values are calculated only for dataset 2.}
 \end{deluxetable*}
We used data from the Atmospheric Imaging Assembly (AIA) on board of the Solar Dynamic Observatory (SDO)
\citep{Lemen12}, which
monitors the Sun with 0.6" spatial and 12 sec temporal resolution  \citep{Boerner2012}. The SDO/AIA data are
retrieved, processed and analyzed using standard procedures with the SolarSoft (SSW) package. Our particular
interest goes to BPs and CJs situated within and at the edges of the CHs. The coronal holes have been chosen from
different
areas
of both hemispheres, excluding regions very close to limbs, during the period from 1 December 2015 till 1 May
2016.
In total, we investigated 23 (spatiotemporarily independent) CJs using SDO/AIA 193 {\AA} channel images. Among them
were recurrent jets which,
nonetheless, we treated each jet phenomenon as a separate event.

For despiking, bad pixel/cosmic-ray influence correction and flat-fielding, level~1 data has been processed into
level~1.5 using the aia$\_$prep.pro code. From the obtained files, we cut out rectangular
boxes for each
event and create their time series. We applied the rot$\_$xy procedure to exclude the solar rotation effect.
Figure~\ref{Fig1} shows several example snapshots from such an image sequence. The images are created using the AIA 193 
{\AA},304 {\AA} and 171 {\AA} channel data both in combination with HMI photospheric magnetograms. Blue and yellow 
contours indicate positive and negative magnetic field polarities, respectively. In the middle (small yellow and blue 
closed contours) of the BP there is a small bipolar closed magnetic structure anchored in the photosphere and extended 
in the form of a magnetic loop through the base of the solar corona. Besides, it is seen that the footpoints of the 
central loop are surrounded by a system of magnetic field concentration areas (edge of supergranules) in the 
chromospheric network. These magnetic flux concentration regions are closely situated to each other and also embedded in 
the global CH magnetic field of positive polarity (Fig.~\ref{Fig1}). Panel (a) shows the beginning of the precursor for 
each wavelength and for the  magnetogram as well; panel (b) corresponds to the peak of the precursor brightening; panel 
(c) corresponds to the time after the precursor when there is still no signature of the main jet outflow; panel (d) 
demonstrates the moment when the structure is destabilized and the jet-type instability starts and filamentary darkening 
aside the BP loop appears, just like in many similar observations; panel (e) corresponds to the fully developed 
transient jet outflow.

In order to perform a rigorous analysis the observed dynamical processes we created two types of 193 {\AA} intensity
curves in Fig.~\ref{Fig2}. The first type of data (dataset 1) comprises the calculation
of the
mean intensity values of the entire cutout BP-boxes. The second type of data (dataset 2) is created using the average
over
all pixels of modified intensity values cutout BP-boxes obtained through noise deduction (using the methodology
similar to one outlined in \citet{Chandrashekhar2013}). The final values in dataset 2
exclude low-intensity noise containing regions of BP-boxes. Consequently, in dataset 2, the effect of background
noise is removed and all the transient disturbances are more sharply observable. All 23 BPs
and
corresponding jets we investigated have more or less similar morphological and dynamical properties. Examples of
brightness evolution curves are shown in top panels (CJ1) and (CJ2/CJ3/CJ4) of Fig.~\ref{Fig2}.

Despite the fact that the examples of the BPs we studied show the structure and dynamics conventionally standardized in
the related literature \citep{Raouafi2016, Sterling2015}. Our study uncovers the systematic presence of
relatively low amplitude (compared to the jets) quasi-oscillatory dynamic processes before each main jet events (the 
solid-line, black parts of the curves in top panels (CJ1) and (CJ2/CJ3/CJ4) of Fig.~\ref{Fig2}). We argue that 
these
processes can represent precursors of jets (the solid-line, red and blue colored parts of the curves in top 
panels (CJ1) and (CJ2/CJ3/CJ4) of Fig.~\ref{Fig2}). Further, we plot these precursor parts zoomed in and detrended for 
an isolated single (top panel (CJ1)) and for a recurrent jet event (panels (CJ2), (CJ3), (CJ4)). It is 
apparent that regardless of the type of the event (single or recurrent) the quasi-oscillatory variation of the mean 
intensity of precursors  is systematically observed.

\section{Analysis of the results}
\begin{figure*}
\begin{center}
\includegraphics[scale=0.95]{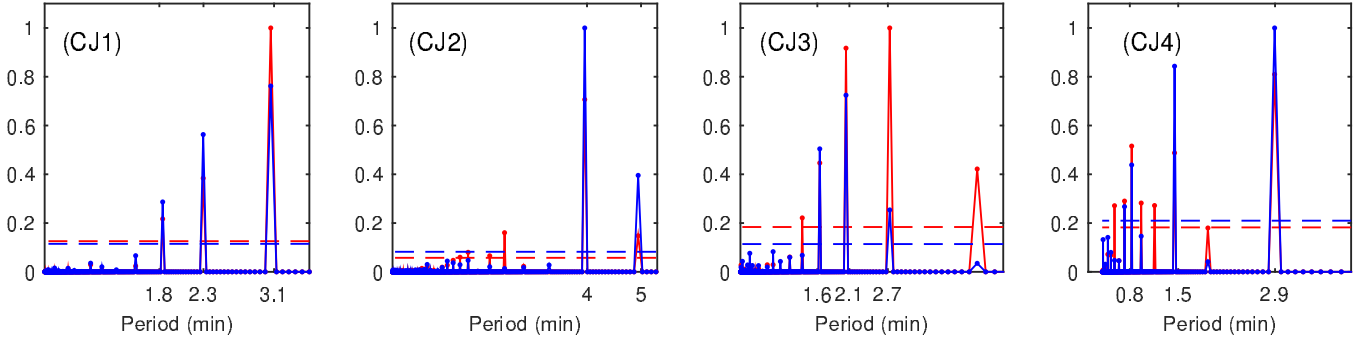}
\end{center}
\caption{Set of calculated FFT periodograms accordingly corresponding to cases shown in bottom panels (CJ1), 
(CJ2), (CJ3), (CJ4) of Fig.~\ref{Fig2}. The coloring is same as in Fig.~\ref{Fig2} and horizontal blue and red dashed 
lines represent $95 \%$ confidence levels for the dataset 2 and dataset 1, respectively. The spectral powers are 
normalized on the maximum power pick shown in each panel.}
\label{Fig3}
\end{figure*}

Our first finding is that in the absolute
majority of instances (20 out of 23) we detect a characteristic brightening of the observed BP preceding, by a few
minutes,
the main jet outflows. We consider these processes as CJ precursors (as shown in Fig.~\ref{Fig2}). In this paper we show
only illustrative examples of single and recurrent jet events and the full analysis all 23 jets will be given in the
form of
\url{http://solspanet.eu/solspanet/static/Supplementary_Data_Bagashvili_1001.pdf}
on our
project web site.

Figure~\ref{Fig2} shows only illustrative examples of single (top panel (CJ1)) and recurrent (panel 
(CJ2/CJ3/CJ4)) jet events
(out of all 23 investigated cases). The enhancement of the BP intensity is visible several minutes before the CJ
ejection in the case of both dataset 1 and dataset 2 mean intensity curves. In both cases of single and recurrent jets 
the precursors are systematically observed before each plasma ejection. In top panel (CJ1) Fig.~\ref{Fig2} 
existence of the precursor of the single jet event is evident (2015-12-09 17:18 UT). Besides, the set of recurrent 
events in the panel (CJ2/CJ3/CJ4) of Fig.~\ref{Fig2} took place within the time interval from 30-12-2015 23:01
UT to 31-12-2015 00:31 UT and included three consecutive plasma ejections, each of them had corresponding precursors.
Finally, for the 23 investigated events a precursor was found in 20 cases. The detailed catalog of the investigated CJ
parameters is presented in Table \ref{tab:mathmode}.

We investigated the statistical distributions of the precursor and the CJ individual parameters for both dataset 1 and 
dataset 2. We organized the parameters as follows (Table~\ref{tab:parameter_means}): (i) The precursor ignition duration 
 ($\tau_{PI}$), representing the half-width of the Gaussian fit to the first disturbance in the precursor. The 
$\tau_{PI}$ contains the information about the temporal properties of the precursor sources which enables us to judge 
the particular periods of the initially ignited disturbances. It should be noticed that some precursors consist of a 
single peak disturbance and in such cases the ignition and total duration of the precursors are co-measurable. In other
cases, however, the precursors include an oscillatory-like behavior, and in this case, the total length of the process 
is significantly larger than the ignition (see Table~\ref{tab:mathmode}). (ii) The precursor total evolution time span 
($\tau_{PT}$) showing how long the precursor process lasts before the jet outflow starts. There is another parameter 
also enabling to judge on time spans between the precursors and the CJs, viz.\ (iii) The time intervals between 
precursor and CJ
peaks ($\Delta \tau_{Peaks}$). (iv) CJ durations ($\tau_{CJ}$) calculated in a similar way as the precursor ignition, as 
described above. We also evaluated the interrelation between the precursors and coronal jets by introducing the 
following dimensionless parameters: (i) the ratio of the coronal jet and precursor ignition peak 
intensities ($I_{CJ}/ I_{PI}$) (the parameter manifests the rate of the coupling and the energy pumping from the 
precursor, and maybe some external source, to the main jet) and (ii) their durations ($\tau_{CJ}/ \tau_{PI}$). Finally, 
the ratio of the temporal gap between the precursor ignition and the jet peaks over (iii) the precursor 
ignition ($\Delta \tau_{Peaks}/ \tau_{PI}$) and (iv) jet ($\Delta \tau_{Peaks}/ \tau_{CJ}$) durations.

\section{Discussion and Conclusions}\label{sec:results}
We performed a statistical analysis of the brightness evolution of 23 small-scale jet-like events located within
on-disk CHs.
We studied the corresponding bright point brightness average intensity evolution in time, using high temporal and
spatial
resolution SDO/AIA 193 {\AA} images. We found that the vast majority of coronal jets (CJs) are accompanied by a minor increase
in mean intensity occurring several minutes before each intensive jet ejection. The precursor was identified in
20 out of 23 cases. We consider such enhancements of brightness as precursors of CJs. This result allows us to conclude
that the presence of the precursor is a systematic property of CJs and the absence of the precursor in
3 of the 23 considered cases might be due to either incompleteness of the observational data.  Alternatively,
this can also be caused by the overlap between the precursor ($\tau_{PI}$) and the jet ($\tau_{CJ}$) duration
intervals, which can happen when both are larger than the characteristic time between the peaks ($\Delta \tau_{Peaks}$) 
(zero values in corresponding fields). Despite the fact that there are some studies of the BP brightness evolution, for 
instance by \citet{Pucci2012}, we performed introduction of the notion of CJ precursors.

According to our observation, the average lifetime of all examined $\tau_{PI}$ is 2.77/2.85$\;$min, the mean ratio of CJ 
and precursor ignition peak intensities ($I_{CJ}/I_P$) is 2.18/2.59, the average time between CJ and its precursor 
($\Delta \tau_{Peaks}$) is 10.25/10.13$\;$min, while the duration of the CJs ($\tau_{CJ}$) is 8.52/8.62$\;$min (see 
Table~\ref{tab:parameter_means}) which is in good agreement with other observational indications \cite[see e.g.][and 
references therein]{Schmieder2013}. These parameters allow drawing preliminary conclusions on the observed processes. 
(i) We made an estimation of the mean maximum oscillation period by introducing the parameter $\tau_{PI}$ and also 
trough the FFT analysis. As we see from the table both estimations coincide very well. (ii) The relation between the 
total ($\tau_{PT}$) and ignition ($\tau_{PI}$) durations of the precursor shows that the longest period oscillation 
makes on average 2-3 oscillations before the jet starts which is also justified by the mean values of ratio $\Delta 
\tau_{Peaks}/\tau_{PI}$ and $\tau_{CJ}/\tau_{PI}$. (iii) The value of $\Delta \tau_{Peaks}/\tau_{CJ}$ demonstrates the 
fact that the mean period of the mode that is involved in the jet outflow must be larger than that of the one involved 
in the precursor. (iv) The ratio of $I_{CJ}/I_P$ proves that the amplitude of the jet mode is larger than that of the 
precursor which indicates that perhaps the presence of the wave mode nonequilibrium driving (entropy variations or shear 
flows). This kind of parameter analysis can be continued. However, we stop at this moment as further analysis ultimately 
requires mathematically rigorous modeling which is planned in the near future (see corresponding remarks below).

The key conclusion that can be drawn from our analysis is that we were able to detect quasi-periodical oscillations 
with characteristic periods from sub-minute up to 3-4 min values in the BP brightness which precede the jets. The AIA 
cadence favors such detection. The basic claim what can be made at this stage of pure observational analysis is that 
along with the conventionally accepted scenario of BP evolution through new magnetic flux emergence and its reconnection 
with the initial structure of the BP and the CH, certain MHD oscillatory and wave-like motions can be excited and these 
can take an important place in the observed dynamics. One can even imagine that these quasi-oscillatory phenomena might 
play the role of links between different epoches of the CJ ignition and evolution. However, we do not have at this 
moment rigorous information neither on the nature of this quasi-oscillatory behavior nor on the mentioned possible links 
with the standard evolutionary scenario of the BPs and CJs. We just give a rough estimation of characteristic periods 
by applying a standard FFT routine to the data. A complete understanding of this issues requires further analytical and 
perhaps even numerical modeling of the processes discovered here and such investigations will become a matter of future 
more extensive studies. These are not compatible with the format of the present short letter, which only aims at the 
announcement of the novel observational evidence. In other words, the goal is to establish a notion of the CJ and their 
precursor patterns. However, we can make some general qualitative indications on the observed oscillatory processes and 
their link with some theoretical background. The quasi-oscillatory variations of intensity can be an indication of the 
MHD wave excitation processes due to the system entropy variations \citep{Shergelashvil2007} or density variations 
\citep{Shergelashvili2005, Zaqarashvili2002}. The observed mutual positioning of open and closed magnetic field 
structures, indicates that there is very likely a sharp outflow velocity gradients at the edges between the open and 
closed field line regions. All these conditions suggest a sequence of local magnetic reconnection events that could be 
the source of MHD waves due to impulsive generation or rapid temperature variations \citep{Shergelashvil2007} on the one 
hand, and shear flow driven MHD wave excitation, coupling and dissipation (self-heating mechanism) processes 
\citep{Shergelashvil2006} and explosive type strongly non-adiabatic overreflections in the shear flows 
\citep{Gogoberidze2004}, on the other hand. The results obtained in this letter create a solid ground for further 
studies in this direction.
\acknowledgments
Work was supported by Shota Rustaveli National Science Foundation grants DI-2016-52 and FR17\_609. 
Work of
S.R.B. was supported
under Shota Rustaveli National Science Foundation grants for doctoral students - PhDF2016\_204 and grant for young 
scientists for scientific research internships abroad IG/50/1/16. B.M.S. and M.L.K. acknowledge the support by the 
Austrian Fonds zur Foerderung der Wissenschaftlichen Forschung within the projects P25640-N27, S11606-N16 and Leverhulme 
Trust grant IN-2014-016. The work of T.V.Z. was supported by the Austrian Science Fund (FWF)  under project P30695-N27  
and from the Georgian Shota Rustaveli National Science Foundation project DI-2016-17. M.L.K. additionally acknowledges 
the support of the FWF projects I2939-N27 and P25587-N27, and the grants No.16-52-14006, No.14-29-06036 of the Russian 
Fund for Basic Research. We are thankful to the referee for constructive comments on our manuscript.

\end{document}